\documentclass{tlp}
\usepackage{epsfig}
\usepackage{eepic}
\usepackage{amssymb}
\newenvironment{example}{%
         \begin{ex} \rm%
         }{%
         $\Box$\end{ex}\ignorespaces%
         }
\newtheorem{theorem}{Theorem}

\newtheorem{ex}[theorem]{Example}
\newtheorem{defn}[theorem]{Definition}

\newcommand{\oldtouched}[1]{}

\newcommand{\simparrow}[0]{\Longleftrightarrow}        
\newcommand{\proparrow}[0]{\Longrightarrow}

\def\tuple#1{\langle #1 \rangle}

\newcommand{\pshort}[1]{}

\long\def\short#1{}
\newenvironment{ttline}{\begin{trivlist}\small\tt\item}{\end{trivlist}}
\newenvironment{ttprog}{\begin{trivlist}\small\tt\item\begin{tabbing}}%
        {\end{tabbing}\end{trivlist}}

 \submitted{31 August 2002}
  \revised{2 March 2004, 28 June 2004}
  \accepted{9 August 2004}                                                     

\begin{document}

\renewcommand{\baselinestretch}{0.98}
\normalsize
 
\title[Optimizing Compilation of Constraint Handling Rules in
HAL]{Optimizing Compilation of Constraint Handling Rules in HAL\footnote{A preliminary version of this
paper appeared under the title 
``Optimizing Compilation of Constraint Handling Rules''
        in {\em ICLP 2001},
        Cyprus, November 2001~\cite{halchr}.
}}

\author[C. Holzbaur  \& M. \ Garc\'{\i}a de la Banda \& 
P.J.\ Stuckey  \& G.J. Duck]{
CHRISTIAN HOLZBAUR \\
Dept.~of Medical Cybernetics and Art. Intelligence, 
University of Vienna, Austria 
\and MARIA GARCIA DE LA BANDA \\
  School of Computer Science \& Software Engineering,
  Monash University, Australia 
\and PETER J. STUCKEY, GREGORY J. DUCK \\
 NICTA Victoria Laboratory, \\
 Department of Computer Science \& Software Engineering,
  University of Melbourne, Australia
}

\date{}
\maketitle

\begin{abstract}
  To appear in Theory and Practice of Logic Programming (TPLP).
  In this paper we discuss the optimizing compilation of Constraint
  Handling Rules (CHRs).  CHRs are a multi-headed committed choice
  constraint language, commonly applied for writing incremental
  constraint solvers.  CHRs are usually implemented as a language
  extension that compiles to the underlying language.  In this paper
  we show how we can use different kinds of information in the
  compilation of CHRs in order to obtain access
  efficiency, and a better translation of the CHR rules into the
  underlying language, which in this case is HAL. 
  The kinds of information used include the types,
  modes, determinism, functional dependencies and symmetries of the
  CHR constraints.
  We also show how to analyze CHR programs to determine this information
  about functional dependencies, symmetries and other kinds of information
  supporting optimizations.

\end{abstract}

\section{Introduction}

Constraint handling rules~\cite{fruehwirth:CHRs} (CHRs) are a very
flexible formalism for writing incremental constraint solvers and
other reactive systems.  In effect, the rules define transitions from
one constraint set to an equivalent constraint set. Transitions serve
to simplify constraints and detect satisfiability and
unsatisfiability.  CHRs have been used extensively (see
e.g.~\cite{chrsissue}). Efficient implementations have been
available for many years in 
the languages SICStus Prolog
and ECL$^{i}$PS$^{e}$, and
implementations for other languages are appearing 
such as Java~\cite{jack} and HAL.

In this paper we discuss  how
to improve the compilation of CHRs by using additional information
derived either from declarations 
provided by the user or from the analysis of the
constraint handling rules themselves. The major improvements we
discuss over previous work on CHR compilation~\cite{chrsissue} are:
\begin{itemize}
\item 
general index structures which are specialized for 
the particular joins required in
the CHR execution. Previous CHR compilation was restricted to two kinds of
indexes: simple lists of constraints for given \texttt{Name/Arity} and lists indexed by the variables
involved.  
For ground usage of CHRs this meant that only list indexes are used.
\item 
continuation optimization, where we use matching information from
rules earlier in the execution to avoid matching later rules.
\item 
optimizations that take into account algebraic properties such as functional
dependencies, symmetries and the set semantics of the constraints.
\end{itemize}
We illustrate the advantages of the various optimizations
experimentally on a number of small example programs in the HAL
implementation of CHRs.  We also discuss how the extra information required
by HAL in defining CHRs (that is, type, mode and determinism information) is
used to improve the execution.

In part some of the motivation of this work revolves around a
difference between CHRs in Prolog and in HAL. HAL is a typed language
which does not (presently) support attributed variables.  Prolog
implementations of CHRs rely on the use of attributed variables to
provide efficient indexing into the constraint store.  
Hence, we are critically interested in
determining efficient index structures for storing constraints in the
HAL implementation of CHRs.  An important benefit of using specific
index structures is that CHRs which are completely ground can still be
efficiently indexed.  This is not exploited in the current Prolog
implementations.  As some CHR solvers only use ground constraints this
is an important issue.

The remainder of the paper is organized as follows.  In the next
section we give preliminary definitions, including the operational
semantics of constraint handling rules.  
In Section~\ref{sec:comp} we
go through the basic steps involved in compiling a set of constraint
handling rules, and see how we can make use of properties such
as functional dependencies and symmetry and set semantics in
improving this basic compilation.
In Section~\ref{sec:analysis} we show how we can
improve the compilation of a set of CHRs 
by discovering properties of constraints 
by reasoning about the form of the rules defining them.  
In Section~\ref{sec:fds} we show how to infer the functional
dependencies and symmetry information, used in Section~\ref{sec:comp}, 
from a set of CHRs. 
In Section~\ref{sec:exp} we
give our experimental results illustrating the advantages of the
optimized compilation.  Finally, in Section~\ref{sec:concl} we
conclude.

\section{Constraint Handling Rules and HAL} \label{sec:chr}

Constraint Handling Rules manipulate a global multiset of primitive 
constraints, using multiset rewrite rules which can take three forms 
\begin{eqnarray*}
{\mbox {\it simplification}} &&
[name @]~ c_1, \ldots, c_n 
        ~~\simparrow~~  g ~|~ d_1, \ldots, d_m \\
{\mbox {\it propagation}} &&
[name @]~ c_1, \ldots, c_n 
        ~~\proparrow~~  g ~|~ d_1, \ldots, d_m \\
{\mbox {\it simpagation}} &&
[name @]~ c_1, \ldots, c_l ~\backslash~ c_{l+1}, \ldots, c_n  
        ~~\simparrow~~  g ~|~ d_1, \ldots, d_m
\end{eqnarray*}
where $name$ is an optional rule name,
$c_1, \ldots, c_n$ are CHR constraints, $g$
is a conjunction of constraints from the underlying language, 
and $d_1, \ldots, d_m$ is a conjunction of
CHR constraints and constraints of the underlying language. 
The guard part $g$ is optional. 
If omitted, it is equivalent to $g \equiv true$. 

The simplification rule states that given a constraint multiset
$\{c_1', \dots, c_n'\}$ and substitution $\theta$ 
matching the multiset $\{c_1, \ldots, c_n\}$, 
i.e.~$\{c_1', \dots, c_n'\} = \theta(\{c_1, \ldots, c_n\})$,
where the execution of $\theta(g)$ succeeds, then we can replace
 $\{c_1', \dots, c_n'\}$ by multiset $\theta(\{d_1, \ldots, d_m\})$.
The propagation rule states that, for a matching constraint multiset 
$\{c_1', \ldots, c_n'\}$ where $\theta(g)$ succeeds, we should add 
$\theta(\{d_1, \ldots, d_m\})$.
The simpagation rules states that, given a matching constraint multiset 
$\{c_1', \ldots, c_n'\}$ where $\theta(g)$ succeeds, 
we can replace 
$\{c_{l+1}', \ldots, c_n'\}$ by $\theta(\{d_1, \ldots, d_m\})$.
A \emph{CHR program} is a sequence of CHRs.

More formally the logical interpretation of the rules is as follows.
Let $\bar{x}$ be the variables occurring in $\{c_1, \ldots, c_n\}$,
and $\bar{y}$ (resp.~$\bar{z}$) be the other variables occurring in
the guard $g$ (resp.~rhs $d_1, \ldots, d_m$) of the rule. 
We assume no variables not in $\bar{x}$ 
appear in both the guard and the rhs.\footnote{
This allows us to more easily define the logical reading, we can always
place a CHR in this form, by copying parts of the guard into the right
hand side of the rule and renaming.}
The logical reading is
\begin{eqnarray*}
{\mbox {\it simplification}} &&
\forall \bar{x} (\exists \bar{y}~ g) \rightarrow ( c_1 \wedge \cdots \wedge c_n 
        \leftrightarrow (\exists \bar{z}~ d_1 \wedge \cdots \wedge d_m)) \\
{\mbox {\it propagation}} &&
\forall \bar{x} (\exists \bar{y}~ g) \rightarrow ( c_1 \wedge \cdots \wedge c_n 
        \rightarrow (\exists \bar{z}~d_1 \wedge \cdots \wedge d_m)) \\
{\mbox {\it simpagation}} &&
\forall \bar{x} (\exists \bar{y}~ g) \rightarrow ( c_1 \wedge \cdots \wedge c_n 
        \leftrightarrow (\exists \bar{z}~c_1 \wedge \cdots \wedge c_l \wedge
d_1 \wedge \cdots \wedge d_m)) 
\end{eqnarray*}

The operational semantics of CHRs exhaustively apply rules to the
global multiset of constraints, being careful
not to apply propagation rules twice on the same constraints (to avoid
infinite propagation).  For more details see e.g.~\cite{CHRopsem}.
Although CHRs have a logical reading (see e.g.~\cite{fruehwirth:CHRs}),
and programmers are encouraged to write confluent CHR programs, there are
applications where a predictable order of rule applications is important.
Hence, the textual order of rules in the program is used to resolve
rule applicability conflicts in favor of earlier rules.

The operational semantics is a
transition system on a triple
$\tuple{s,h,t}_v$ of a set of (numbered) CHR constraints
$s$, a conjunction of
Herbrand constraints $h$, and a set of transitions applied, as
well as a sequence of variables $v$.  
The logical reading of $\tuple{s,h,t}_v$ is as 
$\exists \bar{y}(s \wedge h)$ where $\bar{y}$ are the variables in
the tuple not in $v$.  Since the variable component $v$ never changes
we omit it for much of the presentation.

The transitions are defined as follows: Given a rule numbered $a$
and a tuple $\tuple{s,h,t}_v$ 
$$
c_1, \ldots, c_n \proparrow_a g ~|~ d_1, \ldots, d_k, d_{k+1},\ldots, d_m
$$
where $d_1, \ldots, d_k$ are CHR literals and $d_{k+1},\ldots, d_m$
are Herbrand constraints, 
such that there are numbered literals 
$\{c'_{i_1}, \ldots, c'_{i_n}\} \subseteq s$ where 
$\models h \rightarrow \exists \bar{x} c_1 = c_{i_1} \wedge \cdots \wedge
                                       c_n = c_{i_n}
$
and there is no entry $(i_1, \ldots, i_n, a)$ in $t$ then the transition can
be performed to give new state 
$\tuple{s \cup \{d_1, \ldots, d_k\},
h \wedge d_{k+1} \wedge \cdots \wedge d_m, t \cup \{(i_1, \ldots, i_n,
a)\}}_v$ 
where the new literals in the first component are given new id numbers.

The rule for simplification is simpler.
Given a tuple $\tuple{s,h,t}_v$ and a rule
$$
c_1, \ldots, c_n \simparrow g ~|~ d_1, \ldots, d_k, d_{k+1},\ldots, d_m
$$
such that there are literals   
$\{c'_{i_1}, \ldots, c'_{i_n}\} \subseteq s$ where 
$\models h \rightarrow \exists \bar{x} c_1 = c_{i_1} \wedge \cdots \wedge
                                       c_n = c_{i_n}
$
the resulting tuple is 
$\tuple{s \setminus \{c'_{i_1}, \ldots, c'_{i_n}\} \cup \{d_1, \ldots, d_k\},
h \wedge d_{k+1} \wedge \cdots \wedge d_m, t}_v$ .

In this paper we focus on the implementation of CHRs in a programming
language, such as HAL~\cite{hal}, which requires programmers to provide
type, mode and determinism information. A simple example of a HAL CHR
program to compute the greatest common divisor of two positive numbers $a$ and
$b$ (using the goal \texttt{gcd($a$), gcd($b$)}) is given below.
\begin{ttprog}
:- export \=\= pred hanoi2(int, tower, tower, pair(list(move))). \= \kill
:- module gcd.                                                \>\>\> $(L1)$\\

:- import int. \>\>\> $(L2)$\\
:- chr\_constraint gcd/1. \>\>\> $(L3)$ \\
:- export pred gcd(int).           \>\>\> $(L4)$   \\
:- mode gcd(in) is det.       \>\>\> $(L5)$ \\
base @ gcd(0) <=> true.                  \>\>\> $(L6)$ \\
pair @ gcd(N) \verb+\+ gcd(M) <=> M >= N | gcd(M-N). \>\>\> $(L7)$
\end{ttprog}
The first line $(L1)$ states that the file defines the module {\tt
  gcd}. Line $(L2)$ imports the standard library module {\tt int}
which provides (ground) arithmetic and comparison predicates for the
type \texttt{int}.  
Line $(L3)$ declares the predicate \texttt{gcd/1} to be implemented
by CHRs.
Line $(L4)$ exports the 
CHR constraint {\tt gcd/1} which has one argument, an \texttt{int}.
This is the \emph{type} declaration for {\tt gcd/1}.  Line $(L5)$ is
an example of a \emph{mode of usage} declaration.  The CHR constraint
\texttt{gcd/1}'s first argument has mode \texttt{in} meaning that it
will be fixed (ground) when called.  The second part of the
declaration ``\texttt{is det}'' is a determinism statement. It
indicates that \texttt{gcd/1} always succeeds exactly once (for each
separate call).  For more details on types, modes and determinism
see~\cite{hal,mercury}.

Lines $(L6)$ and $(L7)$ are the 2 CHRs defining the {\tt gcd/1}
constraint.  The first rule is a simplification rule. It states that a
constraint of the form \texttt{gcd(0)} should be removed from the
constraint store to ensure termination.  The second rule is a simpagation rule. It states
that given two different \texttt{gcd/1} constraints in the store, such
that one \texttt{gcd(M)} has a greater argument than the other
\texttt{gcd(N)} we should remove the larger (the one after the
\verb+\+), and add a new \texttt{gcd/1} constraint with argument
\texttt{M-N}.  Note that \texttt{M-N} is the result of the subtraction of
integer \texttt{N} from \texttt{M}, not the term \texttt{-(M,N)} that would
be created in Prolog. Together these rules mimic
Euclid's algorithm.

The requirement of the HAL compiler to always have correct mode 
information means that CHR constraints can only have
declared modes that do not change the instantiation state of their
arguments,\footnote{They may actually change the instantiation state but this
  cannot be made visible to the mode system.} since the compiler will
be unable to statically determine when rules fire. 
Hence for example legitimate modes are \texttt{in}, which means
the argument is fixed at call time and return time, 
and \texttt{oo}, which means that the argument is initialized
at call time, but nothing further is known, and similarly at return time.
The same restriction applies to dynamically scheduled goals in HAL
(see~\cite{hal}).

\section{Optimizing the basic compilation of CHRs}\label{sec:comp}

Essentially, the execution of CHRs is as follows. Every time a new
constraint (the \emph{active constraint}) is placed in the store, we
search for a rule that can fire given this new constraint, i.e., a
rule for which there is now a set of constraints that matches its
left hand side. The first such rule (in the textual order they appear
in the program) is fired.

Given this scheme, 
the bulk of the execution time for a CHR
$$
c_1, \ldots, c_l [,\backslash] c_{l+1}, \ldots, c_n 
        ~~\begin{array}{l}\simparrow\\\proparrow\end{array}~~  g ~|~ d_1, \ldots, d_m 
$$
is spent in determining partner constraints
$c'_1, \ldots, c'_{i-1}, c'_{i+1}, \ldots, c'_n$
for an active constraint $c'_i$ to 
match the left hand side of the CHR.  Hence, for each rule and
each occurrence of a constraint, we are interested in generating
efficient code for searching for partners that will cause the rule to
fire.  We will then link the code for each instance of
a constraint together to form the entire program for the constraint. 
In this section, when applicable, we will show how different kinds
of compile-time information can be used to improve the resulting code
in the HAL version of CHRs.

\subsection{Join Ordering}

The left hand side of a rule together with the guard defines a
multi-way join with selections (the guard) that could be processed in
many possible ways, starting from the active constraint.  This problem
has been extensively addressed in the database literature.  However,
most of this work is not applicable since in the database context they
assume the existence of information on cardinality of relations
(number of stored constraints) and selectivity of various attributes.
Since we are dealing with a programming language we have no access to
such information, nor reasonable approximations.  Another important
difference is that, often, we are only looking for the first possible
join partner, rather than all.  In the SICStus CHR version, the
calculation of partner constraints is performed in textual order and
guards are evaluated once all partners have been identified.
In HAL we determine a best join order and guard scheduling
using, in particular, mode information.

Since we have no cardinality or selectivity information we will select
a join ordering by using the 
number of unknown attributes in the join
to estimate its cost. 
Functional dependencies are used to improve this estimate, 
by eliminating unknown attributes from consideration 
that are functionally defined by
known attributes.

Functional dependencies are represented as $p(\bar{x}) :: S \leadsto x$
where $S \cup \{x\} \subseteq \bar{x}$ meaning that for constraint $p$
fixing all the variables in $S$ means there is at most one solution to the
variable $x$.
The function \textsf{fdclose}($Fixed$,$FDs$) 
closes a set of fixed variables $Fixed$
under the finite set of functional dependencies $FDs$. 
\textsf{fdclose}($Fixed$,$FDs$) is the least set 
$F \supseteq Fixed$ such that for each 
$(p(\bar{x}) :: S \leadsto x) \in FDs$ such that $S \subseteq F$
then $x \in F$.  Clearly the fixedpoint exists, since the
operation is monotonic.

We assume an initial set $Fixed$ of known
variables (which arises from the active constraint), together with the
set of (as yet unprocessed) partner constraints and guards. The
algorithm \textsf{measure} shown in Figure~\ref{fig:jo}, takes as
inputs the set $Fixed$, the sequence $Partners$ of partner constraints
in a particular order, the set $FDs$ of functional dependencies, and
the set $Guards$ of guards and returns the triple
$(score,Goal,Lookups)$.
 
The $score$ is an ordered pair representing the
cost of the join for the particular order given by the $n$ partner
constraints in $Partners$. 
It is made up of the weighted sum 
$(n-1) w_1 + (n-2) w_2 + \cdots + 1 w_{n-1}$ 
of the costs $w_i$ for each individual join with a partner constraint.  
The weighting encourages the cheapest joins to be earliest.

The cost of joining the $i^{th}$ partner constraint to
pre-join expression (the join of the active constraint 
plus the first $(i-1)$ partners), $w_i$, is defined as
the pair $(u,f)$: 
$u$ is the number of arguments in the new partner which
are unfixed before the join; and $f$ is the negative of
the number of arguments which are fixed in the pre-join expression.  
The motivation for this costing is based on the
worst case size of the join, 
assuming each argument ranges over a domain
of the same size $s$.
In this case the number of join partners (tuples) in the partner
constraint for each set of values for the
pre-join expression is $s^u$, 
and there are $s^{m-f}$ tuples
in the pre-join expression (where $m$ is the total number
of variables in the pre-join expression). 
The total number of tuples after the $i^{th}$ partner
is joined are thus $s^{m-f+u}$.
The numbers hence represent the exponents of the join size,
a kind of ``degrees of freedom'' measurement.
The sum of the first components $u$ gives the total size of the
join. The role of the second component is to prefer 
orderings which keep the intermediate results smaller.

We also take into account
the selectivity of the guards we can schedule directly
after the new partner.
This is achieved via the $selectivity(Guards)$ function
which returns the sum of the selectivities of the guards $Guards$.
The selectivity of a equational 
guard $X=Y$ is $1$ provided $X$ and $Y$ are both fixed, 
otherwise the selectivity is $0$.
An equation with both $X$ and $Y$ fixed immediately eliminates one degree
of freedom (reduces the number of tuples by $1/s$), hence
the selectivity of 1. 
When one variable is not fixed, the guard does not remove any answers.
For simplicity, the selectivity of other guards is 
considered to be $0.5$ (as motivation, the constraint $X > Y$ 
where $X$ and $Y$ can be considered to remove 0.5 degrees of freedom).
The role of selectivity is to encourage the early scheduling
of guards which are likely to fail.

$Goal$ gives the ordering of partner
constraints and guards (with guards scheduled as early as possible).
Finally, $Lookups$ gives the queries. Queries will be made from partner
constraints, where a variable name indicates a fixed value, and an
underscore (\_) indicates an unfixed value.  For example,
query \texttt{p(X,\_,X,Y,\_)} indicates a search for
\texttt{p/5} constraints with a given value in the first, third, and
fourth argument positions, the values in the first and third position
being the same.

\begin{figure}[t]
\small
\begin{tabbing}
xx \= xx \= xx \= xx \= xx \= xx \= xx \= xx \=\kill 
\textsf{measure}($Fixed$,$Partners$,$FDs$,$Guards$) \\
\> $Lookups$ := $\emptyset$; $score$ := $(0,0)$; $sum$ := $(0,0)$  \\
\> $Goal$ := \textsf{schedule\_guards}($Fixed$, $Guards$) \\
\> $Guards$ := $Guards \setminus Goal$ \\
\> \textbf{while} $true$ \\
\> \> \textbf{if} $Partners = \emptyset$ \textbf{return} $(score, Goal, Lookups)$ \\
\> \> \textbf{let} $Partners \equiv  p(\bar{x}), Partners1$ \\
\> \> $Partners$ := $Partners1$ \\
\> \> $FD_p$ := $\{ p(\bar{x}) :: fd \in FDs \}$ \\ 
\> \> $Fixed_p$ := $\textsf{fdclose}(Fixed,FD_p)$ \\
\> \> $\bar{f}$ := $\bar{x} \cap Fixed_p$  \\
\> \> $Fixed$ := $Fixed \cup \bar{x}$ \\
\> \> $GsEarly$ := \textsf{schedule\_guards}($Fixed$, $Guards$) \\
\> \> $cost$ := $(max(|\bar{x} \setminus \bar{f}|-selectivity(GsEarly),0), 
                -|\bar{f}|-selectivity(GsEarly))$\\
\> \> $score$ := $score + sum + cost$; $sum$ := $sum$ + $cost$ \\
\> \> $Lookups$ := $Lookups \cup \{ p( (x_i \in \bar{f} ~?~ x_i ~:~ \_ ) ~|~ x_i
\in \bar{x}) \}$ \\
\> \> $Goal$ := $Goal, p(\bar{x}), GsEarly$; \\
\> \> $Guards$ := $Guards \setminus GsEarly$ \\
\> \textbf{endwhile} \\
\> \textbf{return} ($score, Goal, Lookups$) \\
\\
\textsf{schedule\_guards}($F$,$G$) \\
\> $S$ := $\emptyset$ \\
\> \textbf{repeat} \\
\> \> $G_0$ := $G$ \\
\> \> \textbf{foreach} $g \in G$ \\
\> \> \> \textbf{if} $invars(g) \subseteq F$ \\
\> \> \> \> $S$ := $S, g$ \\
\> \> \> \> $F$ := $F \cup outvars(g)$ \\
\> \> \> \> $G$ := $Gs \setminus \{g\}$ \\
\> \textbf{until} $G_0 = G$ \\
\> \textbf{return} $S$ \\
\end{tabbing}
\caption{Algorithm for evaluating join ordering\label{fig:jo}}
\end{figure}

The function \textsf{schedule\_guards}($F$,$G$) 
returns which guards in $G$ can be scheduled given the
fixed set of variables $F$. 
Here we see the usefulness of mode information which allows us
to schedule guards as early as possible. 
For simplicity, we treat mode information
in the form of two functions: $invars$ and $outvars$ which 
return the set of input and output arguments
of a guard procedure. 
We also assume that each guard has exactly one mode
(it is straightforward to extend the approach to multiple modes
and more complex instantiations).
The \textsf{schedule\_guards} keeps adding  guards to
its output argument while they can be scheduled.

The function \textsf{measure} works as follows:
beginning from an empty goal, we first schedule all possible
guards. We then schedule each of the partner constraints $p(\bar{x})$ in
$Partners$ in the order given, by determining the number
of fixed ($\bar{f}$) and unfixed ($\bar{x} \setminus \bar{f}$) variables
in the partner, and the selectivity of any guards that can be scheduled
immediately afterwards. With this we calculate the $cost$ pair
for the join which is added into the $score$.
The $Goal$ is updated to add the join $p(\bar{x})$ followed by the
guards that can be scheduled after it. 
When all partner joins are calculated the function returns.

\begin{example}\label{ex:jo}
Consider the compilation of the rule:
\begin{ttline}
p(X,Y), q(Y,Z,T,U), flag, r(X,X,U) \verb+\+ s(W) ==> W = U + 1, linear(Z) | p(Z,W).  
\end{ttline}
for active constraint \texttt{p(X,Y)} and $Fixed =\{X,Y\}$. The individual 
costs calculated for each join in the
left-to-right partner order illustrated in the rule are 
  $(2.5,-1.5)$, $(0,0)$, $(0,-2)$, $(0,-1)$ 
giving a total cost of
$(10,-11)$ together with goal
\begin{ttline}
q(Y,Z,T,U), W = U + 1, linear(Z), flag, r(X,X,U), s(W)
\end{ttline}
and lookups \texttt{q(Y,\_,\_,\_)}, \texttt{flag}, 
\texttt{r(X,X,U)}, \texttt{s(W)}.
The best order has total cost $(4.5,-7.5)$ 
resulting in goal
\begin{ttline}
flag, r(X,X,U), W = U + 1, s(W), q(Y,Z,T,U), linear(Z) 
\end{ttline}
and lookups \texttt{flag}, \texttt{r(X,X,\_)}, 
\texttt{s(W)}, \texttt{q(Y,\_,\_,U)}.

For active constraint \texttt{q(Y,Z,T,U)}, the 
best order has total cost $(2,-8)$ resulting in goal 
\begin{ttline}
W = U + 1, linear(Z), s(W), flag, p(X,Y), r(X,X,U)  
\end{ttline}
and lookups \texttt{s(W)}, \texttt{flag}, 
\texttt{p(\_,Y)},
\texttt{r(X,X,U)}.
\end{example}

For rules with large-left-hand sides where examining all permutations
is too expensive we can instead
greedily search for a permutation of the partners 
that is likely to be cost effective.
The current HAL implementation uses this method.
In practice, both methods usually find the best ordering because the 
left-hand-sides of CHRs are generally small.

\subsection{Index Selection}

Once join orderings have been selected, we must determine for each
constraint a set of lookups of constraints of that form in the store.  
We then select an index or set of indexes for that constraint that
will efficiently support the lookups required.  Finally, we choose
a data structure to implement each index.
Mode information is crucial to the selection of index data structures.  
If the terms being indexed on are not ground, 
then we cannot use tree indexes since 
variable bindings will change the correct position of
data.\footnote{Currently HAL only supports CHRs with fixed arguments
(although these might be variables from another (non-Herbrand) solver).}

The current SICStus Prolog CHR implementation uses only two index mechanisms:
Constraints for a given \texttt{Functor/Arity} are grouped, 
and variables shared
between heads in a rule {\em index} the constraint store because matching
constraints must correspondingly share a (attributed) variable. In the 
HAL CHR version, we put extra emphasis on indexes for ground data:

The first step in this process is 
\emph{lookup reduction}. Given a set of lookups
for constraint \texttt{p/$k$} we reduce the number of lookups by using information about
properties of \texttt{p/$k$}:
\begin{itemize}
\item lookup generalization: rather than build specialized indexes for
lookups that share variables we simply use more general indexes.
Thus, we replace any lookup \texttt{p}$(v_1, \ldots, v_k)$ 
where $v_i$ and $v_j$ are the same variable by a lookup 
\texttt{p}$(v_1, \ldots, v_{j-1}, v'_j, v_{j+1}, \ldots, v_k)$
where $v'_j$ is a new variable.
Of course, we must add an extra guard $v_i = v_j$ for rules where we use 
generalized lookups.
For example, the lookup \texttt{r(X,X,U)} can use the lookup for 
\texttt{r(X,XX,U)}, followed by the guard \texttt{X = XX}.
\item functional dependency 
reduction: we can use functional dependencies to reduce the
requirement for indexes. 
We can replace any lookup \texttt{p}$(v_1, \ldots, v_k)$ where
there is a functional dependency
$p(x_1, \ldots, x_k) :: \{x_{i_1}, \ldots, x_{i_m}\} \leadsto x_j$
and $v_{i_1}, \ldots, v_{i_m}$ are fixed variables (i.e.~not \_) 
by the lookup \linebreak[4]
\texttt{p}$(v_1, \ldots, v_{j-1}, \_, v_{j+1}, \ldots, v_k)$.
For example, consider the constraint \linebreak[4]
\texttt{bounds(X,L,U)} which stores
the lower \texttt{L} and upper \texttt{U} bounds for a 
constrained integer variable \texttt{X}. Given functional dependency 
$bounds(X,L,U) :: X \leadsto L$,
the lookup \texttt{bounds(X,L,\_)} can be replaced by \texttt{bounds(X,\_,\_)}.
\item symmetry reduction: if \texttt{p/$k$} is symmetric on
arguments $i$ and $j$ we have two symmetric lookups
\texttt{p}$(v_1, \ldots, v_i, \ldots, v_j, \ldots, v_k)$
and \texttt{p}$(v'_1, \ldots, v'_i, \ldots, v'_j, \ldots, v'_k)$
where $v_l = v'_l$  for $1 \leq l \leq k, l \neq i, l \neq j$
and $v_i = v'_j$ and $v_j = v'_i$ then remove one of the symmetric lookups.
 For example, if \texttt{eq/2} is symmetric 
the lookup \texttt{eq(\_,Y)} can use the index for \texttt{eq(X,\_)}.
\end{itemize}

We discuss how we generate functional dependency and symmetry information in
Section~\ref{sec:fds}.  We can now choose the data structures for the
indexes that support the remaining lookups.

Normally, the index will return an iterator which iterates through the
multiset of constraints that match the lookup.  Conceptually, each
index thus returns a list iterator of constraints matching the
lookup.
We can use functional dependencies to determine when this multiset can
have at most one element. This is the case
for a lookup \texttt{p}$(v_1, \ldots, v_k)$ with fixed variables
$v_{i_1}, \ldots, v_{i_m}$ such that \textsf{fdclose}($\{x_{i_1}, \ldots,
x_{i_m}\}, FDp) \supseteq \{x_1, \ldots, x_k\}$ where $FDp$ are the
functional dependencies for \texttt{p/$k$}, since in this case
the functional dependencies ensure that for fixed $v_{i_1}, \ldots, v_{i_m}$
there can be at most one tuple in the constraint.
For example, the lookup \texttt{bounds(X,\_,\_)} returns at most one
constraint given the functional dependencies:
$bounds(X,L,U) :: X \leadsto L$ and $bounds(X,L,U) :: X \leadsto U$.

Since, in general,
we may need to store multiple copies of identical constraints 
(CHR rules accept multisets rather than sets of constraints)
each constraint needs to be stored with a unique identifier, called the
\emph{constraint number}.  Code for the constraint will generate a new
identifier for each new active constraint.
Constraints that cannot have multiple copies stored at once are said to have
\emph{set semantics} (see section \ref{sec:set}).
In this case constraint numbers are not strictly necessary.

Each index for \texttt{p}$(v_1, \ldots, v_k)$, where say
the fixed variables are $v_{i_1}, \ldots, v_{i_m}$,  
needs to support the following operations:
\begin{ttprog}
:- pred p\_index\_init. \\
:- mode p\_index\_init is det. \\
\\
:- pred p\_index\_insert(arg1, \ldots, argk, constraint\_num). \\
:- mode p\_index\_insert(in, \ldots, in, in) is det. \\
\\
:- pred p\_index\_delete(arg1, \ldots, argk, constraint\_num). \\
:- mode p\_index\_delete(in, \ldots, in, in) is det. \\
\\
:- pred p\_index\_init\_iterator(argi1, \ldots, argim, iterator). \\
:- mode p\_index\_init\_iterator(in, \ldots, in, out) is det. \\
\end{ttprog}
for initializing a new index, inserting and deleting constraints
from the index and returning an iterator over the index for
a given lookup. Note that the constraint number is an important
extra argument for index manipulation. In HAL indexes are stored
in global variables, which are destructively updated for
initialization, deletions and insertions.   
The compiler generates code for the predicates
\texttt{p\_insert\_constraint} and \texttt{p\_delete\_constraint}
which insert and delete the constraint \texttt{p} from
each of the indexes in which it is involved.

The current implementation supports three kinds of index structures:
\begin{itemize}
        \item A \texttt{yesno} global variable
        \item A balanced 234 tree
        \item An unsorted list (the default)
\end{itemize}
By far the simplest 
index structure is a \texttt{yesno} global variable, which 
can have
two states: a \texttt{no} state (meaning nothing is currently stored) or a
\texttt{yes(C)} state, where $C$ is the only constraint currently in the
store.
The compiler will generate a \texttt{yesno} index structure
whenever it detects that it is not possible for 
multiple\footnote{These constraints do not have to be identical.} 
\texttt{p/}$k$ constraints to exist in the store at once.
This is the case whenever constraint \texttt{p/}$k$ has set semantics 
(no identical copies) and has the functional dependencies 
$p(\bar{x}) :: \emptyset \leadsto x_i$ for $x_i \in \bar{x}$ (all copies
must
be 
identical).
An example is the constraint \texttt{gcd/1} from the
\texttt{gcd} example program in Section~\ref{sec:chr}.
Here the rule
\begin{ttline}
gcd(N) \verb+\+ gcd(M) <=> M >= N | gcd(M-N).
\end{ttline}
ensures one of the two 
\texttt{gcd/2} constraints (one must be active) will be deleted.
Therefore only one can ever actually be in the store at once,
hence a \texttt{yesno} index structure may be used.

If constraint \texttt{p/}$k$ has set semantics and functional dependencies of 
the form
$p(x_1, \ldots, x_i, x_{i+1}, \ldots, x_k) :: \{x_1, \ldots, x_i\} \leadsto x_j$
for all $i < j \leq k$ then the compiler will generate a balanced 234 tree 
index structure.
In this case the constraint \texttt{p/}$k$ can be thought of as defining a 
function from the key
$(x_1, \ldots, x_i)$ to a value $(x_{i+1}, \ldots, x_k)$.
For example, the constraint \texttt{bounds(X,L,U)} from the 
program in Figure~\ref{fig:interval} in
Appendix A has the functional
dependencies $bounds(X,L,U) :: X \leadsto L$ and
$bounds(X,L,U) :: X \leadsto U$,
hence the compiler builds a 234 tree index structure with \texttt{X} as the
key, and the tuple \texttt{(L,U)} as the value.
In addition, 
if the constraint \texttt{p/}$k$ has set semantics, but has no functional
dependency, then we can still use a tree index 
by treating the entire constraint as the key.
For example, the constraint \texttt{X != Y} from the 
\texttt{interval} program in Appendix A
has set semantics, 
thus we can use a tree structure with \texttt{(X,Y)} as the
key, and the empty set $\emptyset$ as the value.

The big advantage of tree structures is $O(\log(n))$ lookups whenever the
key is fixed, compared with $O(n)$ lookups for unsorted lists.
Even if the key is only partially fixed there is still a potential for 
considerable benefit.
Suppose that $(X,Y)$ is the key, then all keys of the form $(X,\_)$ will group
together in the tree index because of HAL's default ground term ordering.
As a result, we can still do fast 
searches by pruning large sections of the tree
that are not of interest.
The same is not true for keys of the form $(\_,Y)$ but we can sometimes use
symmetric reduction to do the faster $(Y,\_)$ lookup instead.

\begin{example}
The CHR constraint \texttt{!=/2} defined by rules
\begin{ttprog}
neqlower @ \=\= X != Y, bounds(X,VX,VX), \kill 
neqsym @ \>\> X != Y ==> Y != X. \\
neqlower @ \>\> X != Y, bounds(X,VX,VX), bounds(Y,VX,UY) ==> bounds(Y,VX+1,UY).\\
nequpper @ \>\> X != Y, bounds(X,VX,VX), bounds(Y,LY,VX) ==> bounds(Y,LY,VX-1).
\end{ttprog}
has lookups \texttt{!=(X,\_)} and \texttt{!=(\_,Y)}, and \texttt{!=/(X,Y)}
and is known to be symmetric in its two arguments.
We can remove the lookup
\texttt{!=(\_,Y)} in favor of the symmetric \texttt{!=(Y,\_)},
and then use a single balanced tree index for \texttt{!=(X,Y)}
to store \texttt{!=/2} constraints since this can also efficiently
retrieve constraints of the form \texttt{!=(X,\_)}.
\end{example}

The advantage of using a tree is 
lost whenever there is a lookup which is not a prefix of the
index. These lookups can be implemented using 
universal search over the tree, but this 
is particularly bad, since we need to construct a tree iterator, 
which is currently implemented as a tree to 
list conversion with high overhead.
For simplicity, we currently do not use tree indices when at least one lookup
is not a prefix of the key.
Fortunately universal searches against the direction of the functional
dependency are relatively rare in practice, 
and in the future 
implementations of universal searches might do away with the need for
iterators altogether.

The third and final type of index structure is an unsorted list.
The advantage of a list index is fast $O(1)$ insertions, but the disadvantages 
are slow $O(n)$ lookups and deletions.
However, if a constraint \texttt{p/}$k$ is never deleted, and is often involved
in universal searches, then a list is a logical choice for the index structure.

\subsection{Code generation for individual occurrences of active constraints}

Once we have determined the join order for each rule and 
each active
constraint, and the indexes available for each constraint, we are
ready to generate code for each occurrence of the active constraint. 
Two kinds of
searches for partners arise:  A universal
search iterates over all possible partners.  This is required for
propagation rules where the rule fires for each possible
matching partner.  An existential search looks for only the first
possible set of matching partners. 
This is sufficient for simplification rules where
the constraints found will be deleted.

We can split the constraints appearing on the left-hand-side of
any kind of rule into two sets: those that
are deleted by the rule ($Remove$), and those that are not ($Keep$).
The partner search uses universal search behavior, 
up to and including 
the first constraint in the join which appears in $Remove$.
From then on the search is existential.
If the constraint has a functional dependency that
ensures that there can be only one matching solution, we can 
replace universal search by existential search.

For each partner constraint we need to choose an available index 
for finding the matching partners. 
Since we have no selectivity or cardinality information, we
simply choose the index with the largest intersection with the lookup.

\begin{example}
Consider the compilation of the 1st occurrence of
the \texttt{bounds/3} constraint 
in the rule (the fourth occurrence overall 
in the program in Figure~\ref{fig:interval})
\begin{ttprog}
xxxxxxxxxxxxxxxxxxxxxxxx \= \kill
intersect @ bounds(X,L1,U1),bounds(X,L2,U2) <=> \\
\> bounds(X,max(L1,L2),min(U1,U2)). 
\end{ttprog}
Since the active constraint is in $Remove$ the entire search is 
existential.  
The compilation produces the code in Figure~\ref{fig:bounds4}.

\begin{figure}
\begin{ttprog}
xxx \= xxx \= \kill
bounds\_4(X,L1,U1,CN1) :-  \\
\>    (bounds\_index\_exists\_iteration(X,\_,L2,U2,CN2), \\
\> \> CN1 != CN2 -> \\
\>    \> bounds\_remove\_constraint(X,L1,U1,CN1), \\
\>    \> bounds\_remove\_constraint(X,L2,U2,CN2), \\
\>    \> bounds(X,min(L1,L2),max(U1,U2)), \%\% RHS  \\
\>    \> bounds\_4\_succ\_cont(X,L1,U1,CN1) \\
\>    ;  \\
\>    \> bounds\_4\_fail\_cont(X,L1,U1,CN1) \%\% try next rule \\
\>    ). 
\end{ttprog}
\caption{\label{fig:bounds4}Existential search code for the
fourth occurrence of a \texttt{bounds/3} constraint}
\end{figure}

The predicate \texttt{bounds\_index\_exists\_iteration} 
iterates non-deterministically 
through the \texttt{bounds/3} constraints in the store
using the index on the first argument.
In the last 4 arguments
it returns the 3 arguments of \texttt{bounds/3} as well
as a unique constraint number identifying the instance
of the \texttt{bounds/3} constraint.\footnote{The iterator 
need only return the last 2 arguments of \texttt{bounds/3}
and the constraint number, since the first argument must be
known, but the compilation is more straightforward if it always
returns all arguments.}
Note we check that the matching \texttt{bounds/3} constraint
has a different constraint number than the 
active constraint \texttt{CN1 != CN2}. 
The predicate \texttt{bounds\_remove\_constraint} removes the 
\texttt{bounds/3} from the store.
If the matching succeeds, then afterwards we call the
success continuation \texttt{bounds\_4\_succ\_cont} (which will later be
replaced by \texttt{true}),
otherwise we call the failure continuation \texttt{bounds\_4\_fail\_cont}.

The compilation for first occurrence of a \texttt{bounds/3} constraint
in the rule (the second occurrence overall)
\begin{ttline}
redundant @ bounds(X,L1,U1) \verb+\+ bounds(X,L2,U2) <=> L1 >= L2, U1 <= U2 |
true. \\
\end{ttline}
requires universal search for partners since it is not deleted if
the rule succeeds. The compilation produces the code shown in Figure 
\ref{fig:bounds2}.

\begin{figure}
\begin{ttprog}
xxx \= xxx \= \kill
bounds\_2(X,L1,U1,CN1) :-  \\
\>    bounds\_index\_init\_iterator(X,I0), \\
\>    bounds\_2\_forall\_iterate(X,L1,U1,CN1,I0), \\
\>    bounds\_2\_cont(X,L1,U1,CN1). \\
\\
bounds\_2\_forall\_iterate(X,L1,U1,CN1,I0) :- \\
\>    bounds\_iteration\_last(I0), \\
\>    bounds\_3(X,L1,U1,CN1). \\
bounds\_2\_forall\_iterate(X,L1,U1,CN1,I0) :- \\
\>    bounds\_iteration\_next(I0,\_,L2,U2,CN2,I1), \\
\>    (L1 >= L2, U1 <= U2, CN1 != CN2 -> \%\% Guard \\  
\>    \> bounds\_remove\_constraint(X,L2,U2,CN1), \%\% remove matched constraint \\
\>    \> true \%\% RHS \\
\>   ; \\
\> \> true \%\% rule did not apply \\
\> ), \\
\> (bounds\_alive(CN1) -> \\
\> \> bounds\_2\_forall\_iterate(X,L1,U1,CN1,I1) \\
\> ; \\
\> \> true \%\% active has been deleted \\
\> ). \\
\end{ttprog}
\caption{\label{fig:bounds2}Universal search code for the
second occurrence of a \texttt{bounds/3} constraint}
\end{figure}

The predicate \texttt{bounds\_index\_init\_iterator},
returns an iterator of \texttt{bounds/3} 
constraints resulting from looking up the index. 
\texttt{bounds\_iteration\_last} and \linebreak[4]
\texttt{bounds\_iteration\_next} respectively 
succeed if the iterator is finished and return values of the next 
\texttt{bounds/3} (and its constraint number) as well as the new iterator.
After the rule has fired, the predicate \texttt{bounds\_alive} checks
that the active constraint has not been deleted as a consequence of executing
the right-hand-side.
If the active constraint is still alive, then we continue looking for more
matchings.
Note that for universal search we (presently) do not separate fail and 
success continuations.
\end{example}

\begin{example}
Consider the compilation of the 3rd occurrence of
a \texttt{gcd/1} constraint in the program in the introduction
(the second occurrence in $(L7)$) which is to be removed.
Since the active constraint is in 
$Remove$ the entire search is existential.  
The compilation produces the code \texttt{gcd\_3} 
shown in Figure~\ref{fig:gcd}.
\begin{figure}[t]
\small
\begin{minipage}[t]{10cm}
\begin{ttprog}
xxx \= xxx \= \kill
gcd\_3(M,CN1) :-  \\
\>    (gcd\_index\_exists\_iteration(N,CN2), \\
\>     M >= N, CN1 != CN2 -> \%\% guard \\
\>    \> gcd\_delete\_constraint(M,CN1), \\
\>    \> gcd(M-N), \%\% RHS  \\
\>    \> gcd\_3\_succ\_cont(M,CN1) \\
\>    ;  \> gcd\_3\_fail\_cont(M,CN1) ). \\
\\
gcd\_2(N,CN1) :-  \\
\>    gcd\_index\_init\_iterator(I0), \\
\>    gcd\_2\_forall\_iterate(N,CN1,I0), \\
\>    gcd\_2\_cont(N,CN1). \\
\end{ttprog}
\end{minipage}
\begin{minipage}[t]{10cm}
\begin{ttprog}
xxx \= xxx \= \kill
gcd\_2\_forall\_iterate(N,CN1,I0) :- \\
\>    gcd\_iteration\_last(I0), \\
\>    gcd\_insert\_constraint(N,CN1). \\
gcd\_2\_forall\_iterate(N,CN1,I0) :- \\
\>    gcd\_iteration\_next(I0,M,CN2,I1), \\
\>    (M >= N, CN1 != CN2 -> \%\% guard \\ 
\>    \> gcd\_delete\_constraint(M,CN2), \\
\>    \> gcd(M-N) \%\% RHS \\
\>   ; \\
\>    \> true \%\% rule did not apply \\
\>   ), \\
\>   (gcd\_alive(CN1) -> \\
\>   \> gcd\_2\_forall\_iterate(N,CN1,I1) \\
\>   ; \\
\>   \> true \\
\>   ). \\
\end{ttprog}
\end{minipage}
\caption{Code for existential 
partner search and universal partner search.\label{fig:gcd}}
\end{figure}
The predicate \texttt{gcd\_index\_exists\_iteration} 
iterates non-deterministically
through the \texttt{gcd/1} constraints in the store
using the index (on no arguments).
It returns the value of the \texttt{gcd/1} argument 
as well as its constraint number.
Next, the guard is checked. Additionally, we check that the
two \texttt{gcd/1} constraints are in fact different by comparing
their constraint numbers (\texttt{CN1 != CN2}).
If a partner is found, the active constraint is removed from the
store, and the body is called.  Afterwards, the success continuation 
for this occurrence is called. 
If no partner is found the failure continuation is called.

The compilation for second occurrence of a \texttt{gcd/1} constraint
(the first occurrence in $(L7$))
requires universal search for partners. The compilation produces
the code \texttt{gcd\_2} shown in Figure~\ref{fig:gcd}.
The predicate \texttt{gcd\_index\_init\_iterator},
returns an iterator of \texttt{gcd/1} 
constraints resulting from looking up the index. 
Calls to \texttt{gcd\_iteration\_last} and \texttt{gcd\_iteration\_next}
succeed if the iterator is finished 
and return values of the last and next 
\texttt{gcd/1} constraint (and its constraint number) 
as well as the new iterator.
\end{example}

\subsection{Joining the code generated for each constraint occurrence}

After generating the code for each individual occurrence, we must join
it all together in one piece of code.  The occurrences are ordered by
textual occurrence except for simpagation rules where occurrences after
the \verb+\+ symbol are ordered earlier than those before the symbol
(since they will then be deleted, thus reducing the number of
constraints in the store).  Let the order of occurrences be $o_1,
\ldots, o_m$.  The simplest way to join the individual rule code for a
constraint \texttt{p/$k$} is as follows: Code for \texttt{p/$k$} creates
a new constraint number and calls the first occurrence of code
\texttt{p\_$o_1$/$k+1$}.  The fail continuation for
\texttt{p\_$o_j$/$k+1$} is set to \texttt{p\_$o_{j+1}$/$k+1$}.  The
success continuation for \texttt{p\_$o_j$/$k+1$} is also set to
\texttt{p\_$o_{j+1}$/$k+1$} unless the active constraint for this
occurrence is in $Remove$ in which case the success continuation is
$true$, since the active constraint has been deleted.

\begin{example}
For the \texttt{gcd} program the order of the occurrences is $1, 3, 2$. 
The fail continuations simply reflect the order in which the
occurrences are processed: \texttt{gcd\_1} continues to \texttt{gcd\_3}
which continues to \texttt{gcd\_2} which continues to \texttt{true}.
Clearly, the success continuation for occurrences 1 and 3 of \texttt{gcd/1}
are \texttt{true} since the active constraint is deleted. 
The continuation of \texttt{gcd\_2} is \texttt{true} since it is last.
The remaining code for \texttt{gcd/1} is given in
Figure~\ref{fig:rest}.\footnote{Note that 
later compiler passes remove the
overhead of chain rules and empty rules.}
\end{example}
\begin{figure}[t]
\small
\begin{minipage}[t]{10cm}
\begin{ttprog}
xxx \= xxx \= xxx \= xxx \= \kill
gcd(N) :- \\
\> new\_constraint\_number(CN1), \\
\> gcd\_insert\_constraint(N,CN1), \\
\> gcd\_1(N,CN1). \\
gcd\_1(N,CN1) :- \\
\> (N = 0 -> \%\% Guard \\
\> \> gcd\_delete\_constraint(N,CN1), \\
\> \> true, \%\% RHS \\
\> \> gcd\_1\_succ\_cont(N,CN1) \\
\> ; \> gcd\_1\_fail\_cont(N,CN1)). \\
\end{ttprog}
\end{minipage}
\begin{minipage}[t]{10cm}
\begin{ttprog}
xxx \= xxx \= xxx \= xxx \= \kill
\\
gcd\_1\_succ\_cont(\_,\_). \\
gcd\_1\_fail\_cont(N,CN1) :- gcd\_3(N,CN1). \\
\\
gcd\_3\_succ\_cont(\_,\_). \\
gcd\_3\_fail\_cont(N,CN1) :- gcd\_2(N,CN1). \\
\\
gcd\_2\_cont(\_,\_). 
\end{ttprog}
\end{minipage}
\caption{Initial code, code for first occurrence and continuation code for \texttt{gcd}\label{fig:rest}}
\end{figure}

\section{Improving CHR compilation}\label{sec:analysis}

In the previous section we have examined the basics steps for compiling
CHRs taking advantage of type, mode, functional
dependency and symmetries information.
In this section we examine
other kinds of optimizations that can be performed by analysis
of the CHRs.

\subsection{Continuation optimization}

We can improve the simple strategy for joining the code generated for
each occurrence of a constraint by noticing correspondences between
rule matchings for various occurrences.  Suppose we have two
consecutive occurrences with active constraints, partner constraints
and guards given by the triples $(p(\bar{x}), c, g)$ and $(p(\bar{y}),
c', g')$ respectively.  Suppose we can prove that $\models (\bar{x} =
\bar{y} \wedge (\bar{\exists}_{\bar{y}} c' \wedge g')) \rightarrow
\bar{\exists}_{\bar{x}} c \wedge g$ (where $\bar{\exists}_V F$
indicates the existential quantification of $F$ for all its variables
not in set $V$).  Then, anytime the first occurrence fails to match the
second occurrence will also fail to match, since the store has not
changed meanwhile.  Hence, the fail continuation for the first
occurrence can skip over the second occurrence.  

\begin{example}\label{ex:bounds}
Consider the following rules which manipulate
\texttt{bounds(X,L,U)} constraints.
\begin{ttline}
ne  @  bounds(X,L,U) ==> U >= L. \\
red @ bounds(X,L1,U1) \verb+\+ bounds(X,L2,U2) <=> L1 >= L2, U1 <= U2 |
true. \\
int @ bounds(X,L1,U1), bounds(X,L2,U2) <=>
bounds(X,max(L1,L2),min(U1,U2)). 
\end{ttline}
For the 4th and 5th occurrences in rule \texttt{int}
the implication 
$$
(X_4 = X_5 \wedge \exists L2_4, U2_4 bounds(X_4,L2_4,U2_4) ) 
\rightarrow \exists L1_5, U1_5 bounds(X_5, L1_5, U1_5)
$$
(where we use subscripts to 
indicate which is the active occurrence) holds. Hence, the 5th occurrence will never succeed if the 4th fails.
Since if the 4th succeeds the active constraint is deleted, the
5th occurrence can be omitted entirely.
\end{example}

We can similarly improve success continuations. 
If we can prove 
for two consecutive occurrences 
$(p(\bar{x}), c,  g)$ and $(p(\bar{y}), c',  g')$ that 
$
\models \neg \bar{\exists}_{\emptyset}( \bar{x} = \bar{y} \wedge (\bar{\exists}_{\bar{x}} c
\wedge g) \wedge  (\bar{\exists}_{\bar{y}} c'
\wedge g')) 
$
then if the $p(\bar{x})$ occurrence succeeds the $p(\bar{y})$ occurrence
will not. Hence, the
success continuation of $p(\bar{x})$ can skip the 
$p(\bar{y})$ occurrence.
Again, we can use 
whatever form of reasoning we please to prove the
unsatisfiability.
Clearly, this is only of interest when the $p(\bar{x})$ occurrence does not
delete the active constraint.

\begin{example}\label{ex:succcont}
Consider the two occurrences of \texttt{p/2} in the rules:
\begin{ttline}
p(X,Y), q(Y,Y,X,T) ==> X >= Y | ... \\
r(A,B,C), p(C,D) ==> C < D | ...
\end{ttline}
The constraint
$
X = C \wedge Y = D \wedge (\exists T~q(Y,Y,X,T) \wedge X \geq Y) 
\wedge (\exists A,B,C~ r(A,B,C) \wedge C < D)
$
is clearly unsatisfiable and the success continuation
of the first occurrence of \texttt{p/2} can skip the second.
\end{example}

Currently the HAL CHR compiler performs very simple 
fail continuation optimization based on
basic implication reasoning about identical constraints and \texttt{true}.
Furthermore, because of some subtle complications arising from the implementation
of universal searches, the current HAL CHR compiler restricts
continuation optimization to existential searches.
The difficulty stems from deciding if the head of the rule fires or not,
which is information that this optimization relies upon.
For the existential case there is no problem, since matching is already a mere
\texttt{semidet} test.
However a universal search may succeed multiple times, so some additional 
mechanism for recording the number of times a rule fires must be introduced.
One possible solution is thread a counter through the code for the universal
search, and count the number of times the search succeeds.
If the counter is zero after the universal search code exists, then proceed
with the fail continuation, otherwise proceed with the success continuation.
This approach may be implemented in future versions of the
compiler.

\subsection{Late Storage}

The first action in processing a new active constraint is to add it to
the store, so that when it fires, the store has already been
updated.  In practice, this is inefficient since it may quite often be
immediately removed. We can delay the addition of the active
constraint until just before executing a right-hand-side that does not
delete the active constraint, and can affect the store (i.e., may make
use of the CHR constraints in the store).

\begin{example}
Consider the compilation of \texttt{gcd/1}. The 
first and third occurrences delete the active constraint. Thus,
the new \texttt{gcd/1} constraint need
not be stored before they are executed.  
It is only required to be stored just before the
code for the second occurrence.
The call to \texttt{gcd\_insert\_constraint} can be 
moved to  the beginning of \texttt{gcd\_2}, while the calls to
\texttt{gcd\_delete\_constraint} in \texttt{gcd\_1} and \texttt{gcd\_3}
can be removed.
This simplifies the code for \texttt{gcd/1} considerably, as illustrated in
Figure~\ref{fig:gcds}.
\end{example}

\begin{figure}
\begin{ttprog}
xxx \= xxx \= xxx \= xxx \= \kill
gcd(N) :- \\
\> new\_constraint\_number(CN1), \\
\> gcd\_1(N,CN1). \\
gcd\_1(N,CN1) :- \\
\> (N = 0 -> \%\% Guard \\
\> \> true, \%\% RHS \\
\> \> true  \%\% success continuation \\
\> ; \\
\> \> gcd\_3(N,CN1) \%\% fail continuation \\
\> ). \\
gcd\_3(M,CN1) :-  \\
\>    (gcd\_index\_exists\_iteration(N,CN2), \\
\>     M >= N, CN1 != CN2 -> \%\% guard \\
\>    \> gcd\_delete\_constraint(M,CN1), \\
\>    \> gcd(M-N), \%\% RHS  \\
\>    \> true      \%\% success continuation \\
\>    ; \\
\>    \> gcd\_2(M,CN1) \%\% fail continuation \\
\>    ). \\
gcd\_2(N,CN1) :-  \\
\>    gcd\_index\_init\_iterator(I0), \\
\>    gcd\_2\_forall\_iterate(N,CN1,I0). \\
gcd\_2\_forall\_iterate(N,CN1,I0) :- \\
\>    gcd\_iteration\_last(I0), \\
\>    gcd\_insert\_constraint(N,CN1). \\
gcd\_2\_forall\_iterate(N,CN1,I0) :- \\
\>    gcd\_iteration\_next(I0,M,CN2,I1), \\
\>    (M >= N, CN1 != CN2 -> \%\% guard \\
\>    \> gcd\_delete\_constraint(M,CN2), \\
\>    \> gcd\_insert\_constraint(N,CN1), \%\% late insert\\
\>    \> gcd(M-N) \%\% RHS \\
\>   ; \\
\>    \> true \%\% rule did not apply \\
\>   ), \\
\>   (gcd\_alive(CN1) -> \\
\>   \> gcd\_2\_forall\_iterate(N,CN1,I1) \\
\>   ; \\
\>   \> true \\
\>   ). 
\end{ttprog}
\caption{\label{fig:gcds}Simplified code for \texttt{gcd/1} with late
storage}
\end{figure}

The current implementation infers this information by a simple pre-analysis.  
We can consider a rule that does not delete the active constraint as 
\emph{rhs-affects-store} if its right-hand-side calls a CHR constraint, or 
a local predicate which calls CHR constraints (directly or indirectly), or
(to be safe) an external predicate which is not a library predicate.
In future compiler implementations, when CHR constraints are allowed to have 
non-ground arguments, we must also ensure no left-hand-side variables can ever 
be bound by the right-hand-side.
This is because the CHR execution semantics dictate that whenever the 
instantiation state of a constraint changes, we must immediately run that
constraint again as the active.
However since the current implementation only supports ground CHR constraints, 
this issue is not yet relevant.

Good late storage analysis is very important because most of the other analysis
listed in this paper depends on it.
Analysis for detecting set semantics, functional dependencies, never stored
constraints and symmetry all rely on late storage information.

\subsection{Never Stored}

A rule of the form 
$$
c ~~\simparrow~~ d_1, \ldots, d_m
$$ 
where $c$ is a single constraint, always eliminates 
constraints of form $c$ from the store. 
In particular if $c$ is the most general form of the constraint 
$p(x_1, \ldots, x_k),$\footnote{All arguments are pair-wise different
  variables.}
and \texttt{p/$k$} does not need to
be stored because of earlier occurrences of \texttt{p/$k$} in 
rhs-affects-store  rules, 
then we don't need to store this constraint at all.  
The advantage of never-stored information is that any rules involving 
\texttt{p/$k$} will only wake up when $c$ is the active constraint. 
The current implementation searches for instances of never-stored rules and
uses this information to avoid unnecessary joins, and to avoid building 
redundant index structures.

\begin{example}
Consider a \texttt{fixed/1} constraint which 
succeeds if its argument is a variable with equal 
lower and upper bounds, defined by the rules:
\begin{ttprog}
bounds(X,V,V) \verb+\+ fixed(X) <=> true. \\
fixed(X) <=> fail.
\end{ttprog}
Both rules delete the active \texttt{fixed/1} constraint.
Thus, there will never be a \texttt{fixed/1} constraint in the store and
hence an active \texttt{bounds/3} constraint will never match the
rule. Thus, the occurrence of \texttt{bounds/3} in this rule
will not be considered when compiling \texttt{bounds/3}.
\end{example}

\subsection{Set semantics} \label{sec:set}

Although CHRs use a multiset semantics, often the constraints defined by
CHRs have a set semantics.
The current implementation detects two different forms of set semantics.
Either the program rules ensure duplicate copies of constraints are always
deleted or duplicate copies will not affect the behaviour of the program.
The distinction between the two forms affects how the compiler takes advantage 
of this information, but in both cases set semantics allows us to choose more 
efficient index structures.

A constraint \texttt{p/$k$} has \emph{set semantics} if
there is a rule which explicitly removes duplicates of constraints.
That is, if there exists a rule of the form
$$
p(x_1, \ldots, x_k) ~[,\backslash]~ p(y_1, \ldots, y_k) ~~\simparrow~~ g ~|~
d_1, \ldots, d_m
$$
such that $\models x_1 = y_1 \wedge \cdots x_k = y_k \rightarrow
\bar{\exists}_{\bar{x} \cup \bar{y}} g$ which occurs before any rules
require \texttt{p/$k$} to be stored.

\begin{example}\label{ex:bounds-set}
The rule
\begin{ttline}
bounds(X,L1,U1) \verb+\+ bounds(X,L2,U2) <=> L1 >= L2, U2 >= U1 | true.
\end{ttline}
ensures that any new active \texttt{bounds/3}
constraint identical to one already in the store will be
deleted (it also deletes other redundant bounds information).
Since it occurs before any rules requiring \texttt{bounds/3} to be stored
the constraint has set semantics.
\end{example}

A constraint also has set semantics if all rules in which it appears
behave the same even if duplicates are present.
This is a very common case since CHRs are used to build constraint
solvers which (by definition) should treat constraint multisets as sets. 
Thus, a constraint \texttt{p/$k$} also has \emph{set} semantics if
\begin{enumerate}
\item there are no rules which can match two identical 
copies of \texttt{p/$k$} 
\item there are no rules 
that delete a constraint \texttt{p/$k$} without deleting all identical copies.
An exception is when the right-hand-side of such a rule always fails.
\item there are no rules with occurrences of \texttt{p/$k$}
that can generate constraints (on the right-hand-side) which do not have set 
semantics.
\end{enumerate}
The current implementation uses a simple fixpoint analysis which can detect 
such constraints starting from the assumption that all constraints have set 
semantics. In each iteration a constraint which violates one of the rules
above is deleted from the candidate set of those with set semantics.
The iterations proceed until a fixpoint is reached.
 
For constraints \texttt{p/$k$} having this form we can safely 
add a rule of the form
$$
p(x_1, \ldots, x_k) ~\backslash~ p(x_1, \ldots, x_k) ~~\simparrow~~ true.
$$
This will avoid redundant work when duplicate constraints are added.
We can also modify the index structures for this constraint to avoid
the necessity of storing duplicates.

\begin{example}\label{ex:eq}
Consider a constraint \texttt{eq/2} (for equality) defined by the CHR
\begin{ttline}
eq(X,Y),bounds(X,LX,UX),bounds(Y,LY,UY) ==> bounds(Y,LX,UX),bounds(X,LY,UY). 
\end{ttline}
Then, since \texttt{bounds/3} has set semantics,
\texttt{eq/2} also has set semantics. 
If we add the additional rule
\begin{ttline}
eq(X,Y), X != Y <=> fail.
\end{ttline}
then \texttt{eq/2} still has set semantics.
Even though the additional rule might delete one copy only of the constraint
\texttt{eq/2}, 
it does not matter because the rule leads to failure.
\end{example}

\begin{example}
Adding the additional rule which deletes identical copies of constraints can
improve the termination of the program.
Consider the following rules which define symmetry for \texttt{!=} constraints
\begin{ttprog}
neqset @ X != Y \verb+\+ X != Y <=> true. \\
neqsym @ X != Y ==> Y != X.
\end{ttprog}
If we delete the rule \texttt{neqset} then rule \texttt{neqsym} is an infinite
loop for any new \texttt{!=/2} active constraint.
However if \texttt{!=/2} implicitly has set semantics, then we will 
automatically add the rule \texttt{neqset}, hence the program becomes 
terminating.
\end{example}


\section{Determining Functional Dependencies and Symmetries}\label{sec:fds}

In previous sections we have either explained how to determine the
information used for an optimization (as in the case of rules which
are rhs-affects-store) or assumed it was given by the user or inferred
by the compiler in the usual way (as in type, mode and determinism).
The only two exceptions (functional dependencies and symmetries) were
delayed in order not to clutter the explanation of CHR compilation. 
The following two sections examine how to determine these two properties.

\subsection{Functional Dependencies}\label{sec:fd}

Functional dependencies occur
frequently in CHRs since we encode functions using relations.  Suppose
\texttt{p/$k$} need not be stored before occurrences in a rule of the
form
\begin{equation}
\label{fdrule}
p(x_1, \ldots, x_l, y_{l+1}, \ldots, y_k) [,\backslash] 
p(x_1, \ldots, x_l, z_{l+1}, \ldots, z_k) 
~~\simparrow~~ d_1, \ldots, d_m
\end{equation}
where $x_i, 1 \leq i \leq l$ and
$y_i, z_i, l+1 \leq i \leq k$ are distinct variables.
This corresponds to the 
functional dependencies
$p(x_1, \ldots, x_k) :: (x_1, \ldots, x_l) \leadsto x_i, l+1 \leq i \leq k$.
For example, the rule \texttt{int} of Example~\ref{ex:bounds}
illustrates the functional dependencies 
$bounds(X,L,U) :: X \leadsto L$ and $bounds(X,L,U) :: X \leadsto U$.
In addition, rule (\ref{fdrule}) deletes identical copies ensuring
\texttt{p/$k$} has set semantics.
Therefore there is at most one constraint in the store of the form
$p(x_1, \ldots, x_l, \_, \ldots, \_)$ at any time.
Likewise, any constraint \texttt{p/$k$} that has a rule which deletes identical
copies of constraints can also be thought of as having the functional 
dependency
$p(x_1, \ldots, x_k) :: (x_1, \ldots, x_k) \leadsto \emptyset$.

Another common way functional dependencies are expressed in CHRs is by rules 
of the form
$$
p(x_1, \ldots, x_l, y_{l+1}, \ldots, y_k),
p(x_1, \ldots, x_l, z_{l+1}, \ldots, z_k) ~~\proparrow~~ 
y_{l+1}=z_{l+1}, \ldots y_k=z_k
$$
This leads to the same functional dependency as before, 
however it does not lead to set semantics behavior.

We can detect more functional dependencies if we consider multiple
rules of the same kind. For example, the rules
\begin{eqnarray*}
p(x_1, \ldots, x_l, y_{l+1}, \ldots, y_k) [,\backslash]  
p(x_1, \ldots, x_l, z_{l+1}, \ldots, z_k) 
~~\simparrow~~ g_1 | d_1, \ldots, d_m \\
p(x_1, \ldots, x_l, y'_{l+1}, \ldots, y'_k) [,\backslash]  
p(x_1, \ldots, x_l, z'_{l+1}, \ldots, z'_k) 
~~\simparrow~~ g_2 | d'_1, \ldots, d'_{m'}
\end{eqnarray*}
also lead to functional dependencies if 
$\models (\bar{y} = \bar{y}' \wedge \bar{z} = 
\bar{z}' \rightarrow (g_1 \vee g_2)$ is provable.  
However because of the difficulty in solving disjunctions,
the current analysis is limited to the case where $g_1$ and $g_2$ are
primitive integer or real constraints (not conjunctions of other constraints).

\begin{example}\label{ex:gcdall}
The second rule
for \texttt{gcd/1} written twice illustrates the functional dependency
\texttt{gcd}$(N) :: \emptyset \leadsto N$ since
$N = M' \wedge M = N' \rightarrow (M \geq N \vee M' \geq N')$ holds:
\begin{ttprog}
gcd(N) \verb+\+ gcd(M) <=> M >= N | gcd(M - N). \\
gcd(N') \verb+\+ gcd(M') <=> M' >= N' | gcd(M' - N'). 
\end{ttprog}
Making use of this functional
dependency for \texttt{gcd/1} we can use a single global
\texttt{yesno} integer value (\texttt{\$Gcd}) to store the (at most
one) \texttt{gcd/1} constraint, we can replace the forall iteration by
exists iteration, and remove the constraint numbers entirely.  The
resulting code (after unfolding) is
\begin{ttprog}
xxx \= xxx \= xxx \= xxxxxxxxxxxxxxxxxxx \= \kill
gcd(X) :- \\
\> (X = 0 -> true  \>\>\> \%\% occ 1: guard -> rhs\\
\> ; (yes(N) = \$Gcd, X >= N \>\>\>   \%\% occ 3: gcd\_index\_exists\_iteration, guard \\
\>    \> gcd(X-N) \>\>  \%\% occ 3: rhs \\
\>    ; (yes(M) = \$Gcd, M >= X \>\>\> \%\% occ 2: gcd\_forall\_iterate, guard \\ 
\>    \> \$Gcd := yes(X),  \>\> \%\% occ 2: gcd\_insert\_constraint\\
\>    \> gcd(M-X) \>\> \%\% occ 2: rhs \\
\>    ; \$Gcd := yes(X)))). \>\>\> \%\% late insert
\end{ttprog}
\end{example}

\subsection{Symmetry}

Symmetry also occurs reasonably often in CHRs.  There are
multiple ways of detecting symmetries. A rule of the form
$$
p(x_1, x_2, \ldots, x_k) ~~\proparrow~~ p(x_2, x_1, \ldots, x_k) 
$$
that occurs before any rule that requires \texttt{p/$k$}
to be inserted induces a symmetry for constraint $p(x_1, \ldots, x_k)$ 
on $x_1$ and $x_2$, providing that no rule eliminates 
$p(x_1, x_2, \ldots, x_k)$ and not $p(x_2, x_1, \ldots, x_k)$.

\begin{example}
Consider a \texttt{!=/2} constraint defined by the rules:
\begin{ttline}
neqset @  X != Y \verb+\+ X != Y <=> true. \\
neqsym @  X != Y ==> Y != X. \\
neqlower @  X != Y, bounds(X,VX,VX), bounds(Y,VX,UY) ==> bounds(Y,VX+1,UY). \\
nequpper @  X != Y, bounds(X,VX,VX), bounds(Y,LY,VX) ==> bounds(Y,LY,VX-1).
\end{ttline}
the rule \texttt{neqsym @ X != Y => Y != X}
illustrates the symmetry of \texttt{!=/2} w.r.t. $X$ and $Y$, since
in addition no rule deletes a (non-duplicate) \texttt{!=/2} constraint.
\end{example}

A constraint may be symmetric without a specific symmetry adding rule.
The general case is complicated and, for brevity, we simply give examples.

\begin{example}
The rule in Example~\ref{ex:eq} and 
its rewriting with $\{X \mapsto Y, Y \mapsto X\}$ 
are logically equivalent (they are variants illustrated by the
reordering of the rule).
\begin{ttprog}
eq(X,Y),bounds(X,LX,UX),bounds(Y,LY,UY) ==> bounds(Y,LX,UX),bounds(X,LY,UY). \\
eq(Y,X),bounds(Y,LY,UY),bounds(X,LX,UX) ==> bounds(X,LY,UY),bounds(Y,LX,UX).
\end{ttprog}
Hence, since this is the only rule for \texttt{eq/2}, the
\texttt{eq/2} constraint is symmetric.
\end{example}

\begin{example}
The following rules remove redundant inequalities: 
\begin{ttprog}
eq(X,Y) \verb+\+ X <= Y <=> true. \\
eq(Y,X) \verb+\+ X <= Y <=> true. 
\end{ttprog}
They are symmetric for \texttt{eq}($x$,$y$) on $x$ and $y$.
\end{example}

If every rule containing constraint \texttt{p/$k$} is
symmetric on $x_1$ and $x_2$ with another rule then, 
the constraint is symmetric on $x_1$ and $x_2$. 
Hence \texttt{eq}($x$,$y$) is symmetric in $x$ and $y$.


Note that we can take into account symmetries in other constraints
when proving symmetry of rules. 
Hence for example
the additional rule
\begin{ttprog}
X != Y, eq(X,Y) ==> fail.
\end{ttprog}
is symmetric for \texttt{eq}($x$,$y$) on $x$ and $y$
because of the symmetry of \texttt{!=/2}.

Note that we can find more symmetry 
by starting from the assumption that every constraint 
is symmetric on all arguments and
iteratively disproving this.


\section{Experimental Results}\label{sec:exp}

The initial version of the HAL CHR compiler (reported in~\cite{halchr}) 
realized only some of the optimizations discussed herein, 
continuation optimization, simplistic join ordering and
simplistic late storage.
The current version fully implements 
most of the analysis and optimizations
discussed in this paper, including 
\begin{itemize}
        \item early guard scheduling and join ordering;
        \item the discovery of functional dependencies, 
set semantics and symmetries (only the first case of symmetry is detected); 
        \item late storage; and
        \item the building of specialized indexes which 
rely on this information.
\end{itemize}
Analysis is performed in an independent compilation phase, followed by an
optimized code generation phase which builds the index structures amongst other
things.
Further improvements in the analysis phase are possible, such as better
discovery of symmetries (implied from rules) and improved continuation 
optimization.
Further improvements are also possible in the optimized code generation phase,
such as generating multiple indexes for every lookup (currently, the compiler
generates one index per constraint).

The cost of the CHR analysis is small compared with the cost of other tasks
performed by the HAL compiler, such as type and mode analysis.
The discovery of functional dependencies, set semantics and symmetries is
generally very cheap since only a linear pass over the program is required.
Late storage is slightly more expensive since a call graph must be
constructed.
Join ordering is potentially expensive for rules with a large number of 
partners in the head, however the current implementation uses a greedy 
algorithm which is generally much faster.

The currently implementation is a prototype designed to demonstrate how an
optimizing CHR compiler will eventually be realized.
To test the analysis and optimizations implemented by the prototype we compare 
the performance on 3 small programs:

\begin{itemize}
\item \texttt{gcd}  as described in the paper, 
where the query ($a$,$b$) is \texttt{gcd($a$),gcd($b$)}. 
\item \texttt{interval}: 
a simple bounds propagation solver executing N-queens; 
where the query ($a$, $b$)
is for $a$ queens with each constraint added $b$ times 
(usually 1, just here to  illustrate
the possible benefits from set semantics). 
The full code for the bounds propagation solver (module \texttt{interval}) can 
be found in in Appendix \ref{sec:halchr}.
\item \texttt{dfa}: a visual 
parser for deterministic finite automatas (DFAs) 
building the DFA from individual 
graphics elements, e.g. circles, lines and text boxes.
The constraints are all ground, and the compilation
involves a single (indexable) lookup $line(\_,Y)$,
has a single symmetry $line(X,Y) = line(Y,X)$ and no
constraints (except \texttt{line/2}) have set semantics.
In this program the rules are large multi-ways joins, e.g., 
the rule to detect an arrow from one state to another is:
\begin{ttprog}
circle(C1,R1), circle(C2,R2) \verb+\+ \\
line(P1,P2), line(P2,P1), line(P3,P2), line(P2,P3), \\
line(P4,P2), line(P2,P4), text(P5,T) <=> \\
\hspace*{0.5cm} \= point\_on\_circle(P1,C1,R1), 
        point\_on\_circle(P2,C2,R2), \\
\>      midpoint(P1,P2,P12), 
        near(P12,P5) | arrow(P1,P2,T).
\end{ttprog}
Notice that the rule is careful to delete symmetric copies of the constraint 
\texttt{line/2}.
The query $a$ finds a (constant) small DFA (of 10 elements) 
in a large set of $a$ redundant arrows 
(each consisting of three lines and a text box).
\end{itemize}

\begin{table}\small
\caption{Summary of the information extracted by the analysis phase of CHR
compilation}\label{fig:info}
\begin{center}
\begin{tabular}{lllll}
\hline
\hline
Program          & Constraint & FD & Set & Sym \\
\hline
\hline
\texttt{gcd} & \texttt{gcd(X)} & $\emptyset \leadsto X$ & yes & --- \\
\hline
\texttt{interval} & \texttt{bounds(X,L,U)} & $\{X\} \leadsto L,U$ & yes & --- \\
\texttt{interval} & \texttt{eq(X,Y)} & --- & yes & --- \\
\texttt{interval} & \texttt{geq(X,Y)} & --- & yes & --- \\
\texttt{interval} & \texttt{X != Y} & --- & yes & $\{X,Y\}$ \\
\texttt{interval} & \texttt{plus(X,Y,Z)} & --- & yes & --- \\
\hline
\texttt{dfa} & \texttt{line(X,Y)} & --- & yes & $\{X,Y\}$ \\
\hline
\hline
\end{tabular}
\end{center}
\end{table}

A summary of the results from the analysis phase are shown in 
Figure~\ref{fig:info}.
For the \texttt{gcd} program, the analysis infers\footnote{See example 
\ref{ex:gcdall} in Section~\ref{sec:fd}} the functional dependency 
$\emptyset \leadsto X$ and set semantics from the rule
\begin{ttline}
gcd(N) \verb+\+ gcd(M) <=> M >= N | gcd(M-N).
\end{ttline}
The compiler uses this information to build a \texttt{yesno} 
index structure, since the functional dependency combined with set semantics
implies that only one \texttt{gcd($X$)} constraint can ever be in the 
store at one time.

The next program, \texttt{interval}, 
is the most fruitful in terms of information discovered.
The rules
\begin{ttprog}
bounds(X,L1,U1) \verb+\+ bounds(X,L2,U2) <=> L1 >= L2, U2 >= U1 | true. \\
bounds(X,L1,U1), bounds(X,L2,U2) <=> bounds(X,max(L1,L2),min(U1,U2)).
\end{ttprog}
leads to the the discovery of the set semantics of \texttt{bounds/3}
and the functional dependencies $bounds(X,L,U) : X \leadsto L$ 
and $bounds(X,L,U) : X \leadsto U$.
Therefore only one copy of the constraint \texttt{bounds(X,\_,\_)}
can ever be in the store at one time.
The resulting structure for \texttt{bounds/3} is a balanced 234
tree with $(X)$ as the key and $(L,U)$ as the value.

All of the other \texttt{interval} constraints at least have set 
semantics.
Set semantics are inferred for constraints 
that behave the same even if multiple copies are stored.
For example, the only rule involving the constraint \texttt{eq/2} is
\begin{ttprog}
xxxxxxxxxxxxxxxxxxxxxxxxxx \= \kill
equals @ eq(X,Y), bounds(X,LX,UX), bounds(Y,LY,UY) ==> \\
\> bounds(Y,LX,UX), bounds(X,LY,UY).
\end{ttprog}
Since \texttt{bounds/3} has set semantics, then so has \texttt{eq/2}.
The compiler uses this information to eliminate active \texttt{eq/2} 
constraints that already occur in the store.
Thus redundant work is avoided, and potentially the size of indexes is reduced.
In addition symmetry on the constraint \texttt{!=/2} is detected because of 
the symmetric rule
\begin{ttline}
neqsym @ X != Y ==> Y != X.
\end{ttline}
Because of the benefit with symmetric reduction, the compiler will choose a 
balanced 234 tree index for \texttt{!=/2} with $(X,Y)$ as the key.

Finally the \texttt{dfa} program turns out to be the least interesting in
terms of information discovered.
The only constraint with any useful attributes is \texttt{line/2}, which is
symmetric and has set semantics.
Again the compiler generates a balanced 234
tree index (for the same reasons as \texttt{!=/2} 
in the \texttt{interval} program).
All other constraints use the default unsorted list index structure.
However, because of the large size of the heads of rules in the \texttt{dfa}
program, the most important optimization is join ordering and early guard 
scheduling.

\begin{table}\small
\caption{Execution times (ms) for various optimized versions of the \texttt{gcd}
program\label{fig:tablegcd}}
\begin{center}
\begin{tabular}{lrrrrr}
\hline
\hline
Benchmark & Query           & Orig    & +yesno  & +det & Hand \\
\hline                                           
\hline                                           
\texttt{gcd} & (5000000,3)  & 1111    & 976     & 402  & 50   \\
\texttt{gcd} & (10000000,3) & 2314    & 2032    & 803  & 93   \\
\texttt{gcd} & (50000000,3) & 12412   & 11093   & 5095 & 475  \\
\texttt{gcd} & (100000000,3)& 24891   & 22240   & 10270 & 961 \\
\hline
\hline
\end{tabular}
\end{center}
\end{table}

\begin{table}\small
\caption{Execution times (ms) for various optimized versions of the 
\texttt{interval} program\label{fig:tableinterval}}
\begin{center}
\begin{tabular}{lrrrrrr}
\hline
\hline
Benchmark & Query  & Orig  & +tree & +det  & +sym & +eq  \\
\hline
\hline
\texttt{interval} & (12,1) & 389    & 138   & 126   & 67   & 68   \\
\texttt{interval} & (15,1) & 1312   & 382   & 355   & 172  & 169  \\
\texttt{interval} & (20,1) & 6077   & 1602  & 1535  & 693  & 677  \\
\texttt{interval} & (30,1) & 73158  & 11728 & 12537 & 4943 & 4916 \\
\texttt{interval} & (12,2) & 556    & 184   & 167   & 107  & 69   \\
\texttt{interval} & (15,2) & 1824   & 532   & 471   & 283  & 167  \\
\texttt{interval} & (20,2) & 8658   & 2148  & 1984  & 1135 & 669  \\
\texttt{interval} & (30,2) & 110224 & 21950 & 18799 & 8522 & 5071 \\
\hline
\hline
\end{tabular}
\end{center}
\end{table}

\begin{table}\small
\caption{Execution times (ms) for various optimized versions of the \texttt{dfa}
program}\label{fig:tabledfa}
\begin{center}
\begin{tabular}{lrrrrr}
\hline
\hline
Benchmark      & Query   & Prolog  & +join   & +treesym & +det \\
\hline
\hline
\texttt{dfa}   & 20      & 4987    & 30      & 20       & 19  \\
\texttt{dfa}   & 50      & 69070   & 164     & 87       & 86  \\
\texttt{dfa}   & 100     & 532278  & 612     & 271      & 267 \\
\texttt{dfa}   & 200     & too long & 2804  & 1525     & 1536 \\
\texttt{dfa}   & 400     & too long & 13058 & 7401     & 7370 \\
\hline
\hline
\end{tabular}
\end{center}
\end{table}

The results \texttt{gcd}, \texttt{interval} and \texttt{dfa} are shown in 
Table~\ref{fig:tablegcd}, Table~\ref{fig:tableinterval} and Table
~\ref{fig:tabledfa} respectively.
All timings are the average over 20 runs on a 1200MHz AMD Athlon Processor
with 1Gb of RAM running under Debian GNU Linux 3.0 with 
kernel version 2.2.19, and are given in milliseconds. 
Any test taking more than 600000ms (10 minutes) is marked as ``too long''.

For \texttt{gcd} we first give times for the original
output of the compiler $Orig$ (uses a list index).
In the version $+yesno$ the list storage of constraints
is replaced by a $+yesno$ structure (using the functional dependency and set
semantics).
We can see a modest improvement here by just avoiding some overhead.
Note that in $Orig$ the list index for \texttt{gcd/2} never grows more than 
one item in length anyway, so we do not expect a significant improvement by
replacing a singleton list with a \texttt{yesno} structure.
In $+det$ the determinism declarations of the compiled CHR code is altered to 
take into account the functional dependency.
This is a low level optimization in which 
previously ``nondeterministic'' 
lookups can be declared \texttt{semidet} (can succeed at most once).
Without this optimization they are declared 
\texttt{cc\_nondet} which means although they may succeed more than once
we are interested only in the first solution.
This produces faster executable code since
deterministic (including \texttt{semidet}) code 
can be compiled in a simpler way than nondeterministic code.
Finally $Hand$ uses the hand optimized 
implementation of \texttt{gcd}/$1$ shown in Example~\ref{ex:gcdall}.
Here we see a considerable improvement purely from removing all of the overhead
generated by the compiler (such as constraint numbers).
We expect that future implementations of the compiler 
will be able to remove most of this excess overhead.

The second experiment we show is \texttt{interval} in
Table~\ref{fig:tableinterval}.
The original code $Orig$ uses list indexes for all constraints,
version $+tree$ is where the 
list index on \texttt{bounds/3} has been replaced by a
234 tree index (using the functional dependency),
$+det$ where some \texttt{cc\_nondet} searches are 
correctly declared \texttt{semidet},
$+sym$ where the list index on \texttt{!=/2} has been replaced by a 234 tree 
index (because we can take advantage of symmetric reduction),
and $+eq$ where identical copies of set semantic constraints are deleted.
Here we can see a significant improvement when the list index for 
\texttt{bounds/3} is replaced by a 234 tree index.
This is not surprising, since we are replacing $O(n)$ lookups (for lists)
with $O(\log(n))$ lookups (for trees).
Next the $+det$ optimization provides a slight improvement in most cases.
However for some unknown reason the test $(30,1)$ actually becomes slightly 
worse.
Next the $+sym$ lets us take advantage of symmetric reduction, which means we
can choose a 234 tree index for the constraint \texttt{!=/2}.
Again this provides a significant improvement.
Finally the $+eq$ optimization deletes identical copies of constraints before
they run as the active constraint. 
The handling of set semantics is of considerable benefit when
duplicate constraints are actually added, and doesn't add
significant overhead when there are no duplicate constraints, hence it
seems worthwhile.

The final example is the \texttt{dfa} program in Table~\ref{fig:tabledfa}.
The code $Prolog$ has the default join ordering and guard scheduling used by
existing Prolog implementations of CHR compilers.
Recall that this means guards are tested strictly after the join operation,
hence the dreadful performance on a program with large rules, such as
the \texttt{dfa} example.
For the previous examples the default join ordering and the best
join ordering coincide.
Enabling join ordering and 
early guard scheduling ($+join$) produces a massive
improvement in running time (\texttt{dfa} 100 is nearly 2000 times better).
This highlights the importance of this optimization.
Next $+treesym$ turns on 234 tree indexes and symmetric lookup reduction for 
the \texttt{line/2} constraint (without the symmetry lookup reduction, the
compiler will not choose to use a 234 index because of a lookup $line(\_,Y)$).
Once again we get a significant improvement.
Finally in this case the $+det$ optimization seems to produce a 
very slight improvement, if at all.

Finally we remark that it is easy to 
optimize a very poor base implementation of CHRs.  
The HAL base implementation is 
highly efficient. The execution times of the 
$Orig$ or $Prolog$ columns are about an order of magnitude faster
than CHRs in SICStus Prolog. See~\cite{halchr} for more detail.

\section{Conclusion and Future Work}\label{sec:concl}

The core of compiling CHRs is a multi-way join compilation.
But, unlike the usual database case, we have no information on the
cardinality of relations and index selectivity.  We show how
to use type and mode information to compile efficient joins, 
and automatically utilize appropriate indexes for supporting the joins. 
We show how set semantics, 
functional dependencies and symmetries can improve this 
compilation process.
We further investigate how, by analyzing the CHRs themselves we can
find other opportunities for improving compilation, 
as well as determined functional dependencies, symmetries 
and other algebraic features of the CHR constraints.
The prototype HAL CHR compiler which applies these techniques 
produces highly efficient CHR executables.

Almost all of the optimizations considered in this paper 
are not specific to HAL, the optimizations that are not
immediately applicable in a CHR compiler for Prolog are
as follows.
Mode information is not available for guards which means
early guard scheduling may not be as effective, still
assuming all variables are \emph{invars} is safe and will
account for most of the improvement.
The determinism optimization $+det$ in the experiments is
not applicable since determinism declarations
are not supported by Prolog systems.

There is substantial scope for further optimization of CHRs. 
These include: more complicated lookups
(for example range lookups on tree indexes), 
replacing propagation rules by equivalent simplification rules, 
common subexpression elimination,
unfolding of CHRs, and 
determining invariant information for 
stored constraints.
We plan to continue 
improving the HAL CHR compiler to take advantage of these possibilities.


\appendix
\section{Building a constraint solver in HAL using CHRs}\label{sec:halchr}

\begin{figure}
\renewcommand{\baselinestretch}{0.95}
\small
\begin{ttprog}
intersect @ \=\= \kill
:- export \=\> pred hanoi2(int, tower, tower, pair(list(move))). \= \kill
:- module interval.                                                \>\>\> $(L1)$\\
:- import int. \>\>\> $(L2)$\\
:- export\_abstract typedef cint -> f(int). \>\>\> $(L3)$ \\
:- reinst\_old cold = ground. \>\>\> $(L4)$ \\
:- modedef cno -> (new -> cold). \>\>\> $(L5)$ \\
:- modedef coo -> (cold -> cold). \>\>\> $(L6)$ \\
:- VarNum glob\_var int = 0. \>\>\> $(L7)$\\
\\
:- export pred init(cint). \>\>\> $(L8)$ \\
:-     \> mode init(cno) is det.\>\> $(L9)$ \\
init(V) :- V = f(\$VarNum), \$VarNum := \$VarNum + 1, \\
\> (bounds(V,-10000,10000) -> true ; error("not det.")). \\
\\
:- export\_only pred cint=cint. \>\>\> $(L10)$\\
:- \>       mode coo=coo is semidet.\>\> $(L11)$ \\
X = Y :- eq(X,Y). \\
 \\
:- export chrc bounds(cint,int,int). \>\>\> $(L12)$\\
:- \>     mode bounds(coo,in,in) is semidet. \>\> $(L13)$\\
non\_empty@ \>\> bounds(X,L,U) ==> U >= L. \\
redundant@ \>\> bounds(X,L1,U1) \verb+\+ bounds(X,L2,U2) <=> \\
\> ~~~~~~ \= L1 >= L2, U2 >= U1 | true. \\
intersect@ \> bounds(X,L1,U1), bounds(X,L2,U2) <=> \\
\> \> bounds(X,max(L1,L2),min(U1,U2)). \\
\\
:- chrc eq(cint,cint).           \>\>\> $(L14)$   \\
:- mode eq(in,in) is semidet.       \>\>\> $(L15)$ \\
equals @ \> eq(X,Y), bounds(X,LX,UX), bounds(Y,LY,UY) ==> \\
\> \> bounds(Y,LX,UX),bounds(X,LY,UY). \\
\\
:- export chrc cint >= cint. \>\>\> $(L16)$ \\
:- \> mode coo >= coo is semidet. \>\> $(L17)$  \\
geq @ \> X >= Y, bounds(X,LX,UX), bounds(Y,LY,UY) ==> \\
\> \> bounds(Y,LX,UY), bounds(X,LX,UY). \\
\\
:- export chrc cint != cint. \\
:- \> mode coo != coo is semidet. \\
neqset @ \> X != Y \verb+\+ X != Y <=> true. \\
neqsym @ \> X != Y ==> Y != X. \\
neqlower@ \> X != Y,bounds(X,VX,VX),bounds(Y,VX,UY)==>bounds(Y,VX+1,UY). \\
nequpper@ \> X != Y,bounds(X,VX,VX),bounds(Y,LY,VX)==>bounds(Y,LY,VX-1). \\
\\
:- export func cint + cint --> cint. \\ 
:- \> mode coo + coo --> oo is semidet. \\
X + Y --> Z :- plus(X,Y,Z). \\
\\
:- chrc plus(cint,cint,cint). \\
:- mode plus(in,in,in) is semidet. \\
plus@ \=plus(X,Y,Z),bounds(X,LX,UX),bounds(Y,LY,UY),bounds(Z,LZ,UZ)==>\\
\>  bounds(X,LZ-UY,UZ-LY),bounds(Y,LZ-UX,UZ-LX),bounds(Z,LX+LY,UX+UY).
\end{ttprog}
\caption{A simple integer bounds propagation 
solver using CHRs\label{fig:interval}}
\renewcommand{\baselinestretch}{1.00}
\normalsize
\end{figure}

The program in Figure~\ref{fig:interval} defines a simple
bounds propagation solver for integers using constraint handling rules.
From a HAL perspective it is a solver module defining a solver on
the type \texttt{cint}. Line $(L3)$ is the type definition for the new
type \texttt{cint} which is a wrapped integer. The integer is used as a
variable number.  The integer is wrapped so that we have a new type that
we can (re-)define equality for.
 The type is exported abstractly hence its definition is not visible
outside the module, thus restricting operations on \texttt{cint}
to those in this module.   

Line $(L4)$ is a re-instantiation declaration, which is required because
we are going to treat \texttt{cint}s in two ways.  
The \texttt{reinst\_old} declares a new instantiation {\tt cold} (to be
associated with the {\tt cint} type) which is equivalent to ``{\tt old}''
(i.e. a possibly non-ground term) outside the module, and equivalent to 
{\tt ground} inside the module.
We require this because outside the module we treat {\tt cint}s as bounds
propagation solver variables, whereas inside the module they will be
manipulated as wrapped integers (which are \texttt{ground}).
Lines $(L5)-(L6)$ give two common modes of usage for {\tt cint}s.

Variable numbers (for new {\tt cint}s) are kept track of in a global integer 
counter \texttt{VarNum}. This is declared in line $(L7)$ with its type 
{\tt int}, and initial value ($0$).

For any solver type we need to define two predicates \texttt{init/1}
which initializes a new variable and \texttt{=/2} for 
equating two solver variables.
Line $(L8)$ is the predicate declaration for \texttt{init/1}
which is exported.  Its mode is given in line $(L9)$, the mode
\texttt{cno} takes a new object and returns a \texttt{cold}
object (\texttt{old} outside this module, and \texttt{ground} inside this
module). 
Its definition in the next line simply returns the wrapped counter
value, and increments the counter.
The predicate must always succeed exactly once hence its determinism is
\texttt{det}, but to pass determinism checking 
the call to \texttt{bounds/3} is wrapped in an if-then-else (since the
compiler cannot determine that it will not fail).

The \texttt{=/2} predicate is defined in line $(L10)$ 
as \texttt{export\_only}, which makes
it visible outside the module, but not visible inside the module.  This is
to avoid confusion with the equality on the internal view of \texttt{cint}s
which simply treats them as ground terms rather than integer variables.
Its mode definition in line $(L11)$ takes two \texttt{cold} \texttt{cint}s 
as input and returns the same instantiation (the {\tt coo} mode).  It may fail,
so the determinism is \texttt{semidet}.  
Note the definition of \texttt{=/2} is made in terms of the non-exported 
constraint \texttt{eq/2}.

Finally we arrive at our first constraint.  
Line $(L12)$ defines an exported constraint \texttt{bounds/3}
which relates a \texttt{cint} to two \texttt{int}s.
The mode declaration on line $(L13)$ declares that the 
constraint must be invoked with an \texttt{old} \texttt{cint}
and two ground integers.  
The CHR \texttt{non\_empty} is a simple propagation rule. Note the
advantages of a typed language, the \texttt{>=} on the right hand side
is integer comparison, not to be confused with the constraint \texttt{>=}
defined on line $(L16)$.
The CHR \texttt{redundant} removes redundant bounds constraints.
The CHR \texttt{intersect} replaces two \texttt{bounds/3} constraints
on the same variable by one.
Note that \texttt{bounds/3} occurs in many of the rules in the program
not just the two defined immediately below its declaration.

When compiling the module \texttt{interval} we can determine
the functional dependencies: $bounds(X,L,U) :: X \leadsto L$ and
$bounds(X,L,U) :: X \leadsto U$, the symmetries
$eq(X,Y) \equiv eq(Y,X)$,
$neq(X,Y) \equiv neq(Y,X)$ and $plus(X,Y,Z) \equiv plus(Y,X,Z)$
and that each of the CHR constraints has a set semantics.

The lookups required for the program are (after reduction by functional
dependencies):
$bounds(X,\_,\_)$,
$eq(X,_)$, $eq(\_,Y)$, $neq(X,Y)$, $neq(X,_)$,
$neq(\_,Y)$, $plus(X,\_,\_)$, $plus(\_,Y,\_)$ and $plus(\_,\_,Z)$.
Symmetry eliminates the indexes $eq(\_,Y)$, $neq(\_,Y)$ and 
$plus(\_,Y,\_)$.

\end{document}